# The matrix method of representation, analysis and classification of long genetic sequences

Ivan V. Stepanyan, Sergey V. Petoukhov

Mechanical Engineering Research Institute
of the Russian Academy of Sciences, Moscow
e-mail: neurocomp.pro@gmail.com, spetoukhov@gmail.com, petoukhov@imash.ru

**Comment:** Some materials of this article were presented in the "Symmetry Festival-2013" (Delft, Netherlands, August 2-7, 2013, http://symmetry.hu/festival2013.html)

**Abstract.** The article is devoted to a matrix method of comparative analysis of long nucleotide sequences by means of a presentation of each sequence in a form of three digital binary sequences. This method uses biochemical attributes of nucleotides and it also uses a possibility of presentation of every whole set of n-mers in a form of one of members of a Kronecker family of genetic matrices. Due to this method, a long nucleotide sequence can be visually represented as an individual fractal-like mosaic or another regular mosaic of binary type. In contrast to natural nucleotide sequences, artificial random sequences give non-regular patterns. Examples of binary mosaics of long nucleotide sequences are shown, including cases of human chromosomes and penicillins. Interpretation of binary presentations of nucleotide sequences from the point of view of Gray code is also tested. Possible reasons of genetic meaning of Kronecker multiplication of matrices are analyzed. The received results are discussed.

**Content**

## 1. Introduction.

The article is devoted to a matrix method proposed by I. Stepanyan to study long genetic sequences on the base of approaches of "matrix genetics" from works [Petoukhov, 2001, 2008, 2011, 2012a,b, 2013a,b; Petoukhov, He, 2010]. The matrix genetics studies matrix representations of natural ensembles of molecular genetic elements to reveal hidden regularities in cooperative genetic structures and to model inherited biological phenomena, whose features should be agreed with the structural organization of the genetic code for their transferring along a chain of generation.

The notion of "a long nucleotide sequence" usually means a sequence with more then 50000 nucleotides [Prahbu, 1993]. To get more representative visual patterns in a result of application of the described method, longer lengths of nucleotide sequences are preferable.

Known scientific methods for studying nucleotide sequences usually concentrate their attention on those fragments (or n-mers, or n-plets), which exist inside sequences. In contrary to them, our method concentrates on studying those fragments of nucleotide sequences, which are missing in them. In other words, the described meth-od investigates a deficit of different types of n-plets (or-mers) in nucleotide sequences. The authors suppose that this method can be useful for comparative analysis and classification of long genetic sequences and also for deeper understanding of genetic phenomena.

One should also emphasized that this method introduces in the field of molecular genetics and bioinformatics an important theme and notion of binary fractals, which were known in mathematics, physics, informatics and technologies. The theme of binary fractals can be used as a new useful bridge among biological and non-biological fields of sciences and technologies.

Let us remind some approaches to matrix representations of molecular-genetic alphabets from works [Petoukhov, 2001, 2008, 2012a,b; Petoukhov, He, 2010].

## 2. Matrix representations of whole sets of n-plets (or n-mers)

The genetic code system is based on sets or alphabets of n-plets (or n-mers):
- the set of $4^1$ monoplets (A, C, G, T/U);
- the set of $4^2$=16 duplets (AA, AC, AG, AT, ....);
- the set of $4^3$=64 triplets (AAA, AAC, ACA, ACG, ACT, ....);
- etc.

Each whole set of $4^n$ n-plets coincides with the whole set of $4^n$ entries in a $(2^n*2^n)$-matrix, which belongs to the Kronecker family of genetic matrices [A G; C T]$^{(n)}$, where (n) means Kronecker (or tensor) power. Figure 1 shows the first three members of this Kronecker family for n=1, 2, 3. Also Figure 1 shows that - inside such matrix [A G; C T]$^{(n)}$ - each n-plet has its individual binary coordinates (or appropriate coordinates in decimal notation) due to biochemical attributes of n-plets, but this should be explained now specially.

|   | 0 | 1 |
|---|---|---|
| 1 | A (0,1) | G (1,1) |
| 0 | C (0,0) | T (1,0) |

|        | 00 (**0**) | 01 (**1**) | 10 (**2**) | 11 (**3**) |
|--------|-----------|-----------|-----------|-----------|
| 11 (**3**) | AA (00,11) | AG (01,11) | GA (10,11) | GG (11,11) |
| 10 (**2**) | AC (00,10) | AT (01,10) | GC (10,10) | GT (11,10) |
| 01 (**1**) | CA (00,01) | CG (01,01) | TA (10,01) | TG (11,01) |
| 00 (**0**) | CC (00,00) | CT (01,00) | TC (10,00) | TT (11,00) |

|        | 000 (**0**) | 001 (**1**) | 010 (**2**) | 011 (**3**) | 100 (**4**) | 101 (**5**) | 110 (**6**) | 111 (**7**) |
|--------|------------|------------|------------|------------|------------|------------|------------|------------|
| 111 (**7**) | AAA (000,111) | AAG (001,111) | AGA (010,111) | AGG (011,111) | GAA (100,111) | GAG (101,111) | GGA (110,111) | GGG (111,111) |
| 110 (**6**) | AAC (000,110) | AAT (001,110) | AGC (010,110) | AGT (011,110) | GAC (100,110) | GAT (101,110) | GGC (110,110) | GGT (111,110) |
| 101 (**5**) | ACA (000,101) | ACG (001,101) | ATA (010,101) | ATG (011,101) | GCA (100,101) | GCG (101,101) | GTA (110,101) | GTG (111,101) |
| 100 (**4**) | ACC (000,100) | ACT (001,100) | ATC (010,100) | ATT (011,100) | GCC (100,100) | GCT (101,100) | GTC (110,100) | GTT (111,100) |
| 011 (**3**) | CAA (000,011) | CAG (001,011) | CGA (010,011) | CGG (011,011) | TAA (100,011) | TAG (101,011) | TGA (110,011) | TGG (111,011) |
| 010 (**2**) | CAC (000,010) | CAT (001,010) | CGC (010,010) | CGT (011,010) | TAC (100,010) | TAT (101,010) | TGC (110,010) | TGT (111,010) |
| 001 (**1**) | CCA (000,001) | CCG (001,001) | CTA (010,001) | CTG (011,001) | TCA (100,001) | TCG (101,001) | TTA (110,001) | TTG (111,001) |
| 000 (**0**) | CCC (000,000) | CCT (001,000) | CTC (010,000) | CTT (011,000) | TCC (100,000) | TCT (101,000) | TTC (110,000) | TTT (111,000) |

Figure 1. The beginning of the Kronecker family of symbolic genomatrices [A G; C T]$^{(n)}$ for n = 1, 2, 3. Inside each genomatrix [A G; C T]$^{(n)}$, each row and each column has its individual binary numeration due to genetic sub-alphabets (see explanation in text below). Correspondingly each n-plet, which is located on a row-column crossing, has two digital binary coordinates in such matrix. The decimal equivalents of these binary numbers are shown by red color.

The four nitrogenous bases - adenine A, guanine G, cytosine C, thymine T (or uracil U in RNA)) - represent specific poly-nuclear constructions with special biochemical properties. The set of these four constructions is not absolutely heterogeneous, but it bears the substantial symmetric system of distinctive-uniting attributes (or, more precisely, pairs of an "attribute-antiattribute"). This system of pairs of opposite attributes divides the genetic four-letter alphabet into various three pairs of letters by all three possible ways; letters of each such pair are equivalent to each other in accordance with one of these attributes or with its absent.

The system of such attributes divides the genetic four-letter alphabet into various three pairs of letters, which are equivalent from a viewpoint of one of these attributes or its absence: 1) C=T and A=G (according to the binary-opposite attributes: "pyrimidine" or "non-pyrimidine", that is purine); 2) A=C and G=T (according to the binary-opposite attributes "keto" or "amino" [Karlin, Ost, Blaisdell, 1989]); 3) C=G and A=T (according to the attributes: three or two hydrogen bonds are materialized in these complementary pairs). The possibility of such division of the genetic alphabet into three binary sub-alphabets is known from the work [Karlin, Ost, Blaisdell, 1989]. We will utilize these known sub-alphabets by means of the following approach in the field of matrix genetics. We will attach appropriate binary symbols "0" or "1" to each of the genetic letters from the viewpoint of each of these sub-alphabets. Then we will use these binary symbols for binary numbering the columns and the rows of the genetic matrices of the Kronecker family.

Let us mark the mentioned three kinds of binary-opposite attributes by numbers $N = 1, 2, 3$ and let us ascribe to each of four genetic letters the symbol "$0_N$" (the symbol "$1_N$") in case of presence (of absence correspondingly) of the attribute under number "$N$" at this letter. In result we receive the following representation of the genetic four-letter alphabet in the system of its three "binary sub-alphabets to attributes" (Figure 2).

|  | Symbols of a genetic letter from a viewpoint of the binary-opposite attributes | A | G | T/U | C |  |
|---|---|---|---|---|---|---|
| №1 (X) | $0_1$ – pyrimidine; $1_1$ – purine | $1_1$ | $1_1$ | $0_1$ | $0_1$ | 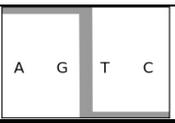 |
| №2 (Y) | $0_2$ – amino; $1_2$ – keto | $0_2$ | $1_2$ | $1_2$ | $0_2$ | 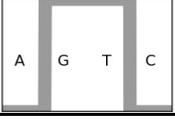 |
| №3 (Z) | $0_3$ – three hydrogen bonds; $1_3$ – two hydrogen bonds | $1_3$ | $0_3$ | $1_3$ | $0_3$ | 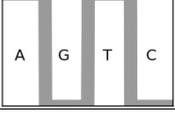 |

Figure 2. Three binary sub-alphabets according to three kinds of binary-opposite attributes in the set of nitrogenous bases C, A, G, T/U. Symbols X, Y, Z in the left column mean names of axes of Cartesian systems of coordinates. Schemes in the right column symbolize graphically each of sub-alphabets.

The table on Figure 2 shows that, on the basis of each kind of the attributes, each of the letters A, C, G, T/U possesses three "faces" or meanings in the three binary sub-alphabets. On the basis of each kind of the attributes, the genetic four-letter alphabet is curtailed into the two-letter alphabet. For example, on the basis of the first kind of binary-opposite attributes we have (instead of the four-letter alphabet) the alphabet from two letters $0_1$ and $1_1$, which one can name "the binary sub-alphabet to the first kind of the binary attributes".

Accordingly, any genetic message as a sequence of the four letters C, A, G, T consists of three parallel and various binary texts or three different sequences of zero and unit (such binary sequences are used at storage and transfer of the information in computers). Each from these parallel binary texts, based on objective biochemical attributes, can provide its own genetic function in organisms.

In view of these three binary sub-alphabets, any nucleotide sequence can be represented as three binary sequences. For example, the sequence ATGGC ... is represented as:

- 10110 ... (in accordance with the first sub-alphabet; its decimal equivalent can be located on the "X" axis of a Cartesian system of coordinates);
- 01110 ... (in accordance with the second sub-alphabet; its decimal equivalent can be located on the "Y" axis of a Cartesian system of coordinates);
- 11000 ... (in accordance with the third sub-alphabet; its decimal equivalent can be located on the "Z" axis of a Cartesian system of coordinates).

For an unambiguous determination of the nucleotide sequence is sufficient to know its binary representations in any two of the three sub-alphabets [Petoukhov, 2001, 2008, 2010]. In particularly, in this example of the sequence ATGGC... , to get its third binary representation 11000 ... (in accordance with the third sub-alphabet) it is enough to summarize two its other representations 10110... and 01110... (received in accordance with the first two sub-alphabets) by means of modulo-2 addition.

In genetic matrices of the Kronecker family (see Figure 1), each row has its individual binary number, which is connected with the fact that all n-plets inside this row have identical binary representation from the point of view of the first sub-alphabets on Figure 2. For example, in the (8x8)-matrix [A G; C T]$^{(3)}$ on Figure 1, the second row has its binary numeration 110 because every of its triplets (AAC, AAT, AGC, AGT, GAC, GAT, GGC, GGT) is a sequence "purine-purine-pyrimidine" that corresponds to binary number 110 from the point of view of the first sub-alphabet on Figure 2. Analogically in genetic matrices of the Kronecker family (see Figure 1), each column has its individual binary number, which is connected with the fact that all n-plets inside this column have identical binary representation from the point of view of the second sub-alphabet on Figure 2. For example, in the (8x8)-matrix [A G; C T]$^{(3)}$ on Figure 1, the third column has its binary numeration 010 because every of its triplet (AGA, AGC, ATA, ATC, CGA, CGC, CTA, CTC) is a sequence "amino-keto-amino" that corresponds to binary number 010 from the point of view of the second sub-alphabet on Figure 2. Respectively, each n-plet, which is located in an appropriate genetic matrix on crossing "column-row", obtains its individual 2-dimensional coordinates on the base of binary numeration of its column and row. For example, the triplet AGC, which is located on crossing of the mentioned column and row (Figure 1), obtains its individual binary coordinates (010, 110) or in decimal notation (2, 6).

Any long nucleotide sequence can be divided into equal pieces of arbitrary length, and a binary record of these fragments can be read in the decimal notation. Then any long nucleotide sequence is represented in the form of three different sequences of decimal numbers, and for its unique identification is sufficient to know its decimal representation in any two sub-alphabets.

If one divides a long nucleotide sequence into equal fragments, whose lengths are equal to "n" (n-mers or n-plets), then each of these fragments is defined by means of its two binary representations (from points of view of the two sub-alphabets) or by means of their equivalents in decimal notations. For example the 5-mer ATGGC is represented as 10110 (in accordance with the first sub-alphabet) and 01110 (in accordance with the second sub-alphabet). Its appropriate decimal meanings are 22 and 14. In such way, this 5-mer ATGGC can be represented not only as the appropriate cell with coordinates (22, 14) inside the genomatrix [A G; C T]$^{(5)}$ but also as the point with decimal coordinates (22, 14) in the orthogonal Cartesian system of coordinates (x,y). Taking into account the chosen connection (Figure 2) between each sub-alphabet and one of axes X, Y, Z of the Cartesian system of coordinates, the following correspondence exists between Kronecker families of genomatrices and 2-dimensional planes (x,y), (x,z) and (y,z) of the Cartesian system:
- the plane (x,y) corresponds to matrices [A G; C T]$^{(n)}$, whose rows and columns are binary numerated from the point of view of the first sub-alphabet and the second sub-alphabet respectively;
- the plane (x,z) corresponds to matrices [G A; C T]$^{(n)}$, whose rows and columns are binary numerated from the point of view of the first sub-alphabet and the third sub-alphabet respectively;
- the plane (y,z) corresponds to matrices [G T; C A]$^{(n)}$, whose rows and columns are binary numerated from the point of view of the second sub-alphabet and the third sub-alphabet respectively.

Taking into account this 2-dimensional representation of each n-plet, one can introduce a notion of Euclidean distance R between any pair of n-plets V($a_1$,$b_1$) and W($a_2$,$b_2$): R = [($a_2$-$a_1$)$^2$ + ($b_2$-$b_1$)$^2$]$^{0.5}$. One can also introduce notions of distance of some other types.

The method, which is described below, uses many variants of a division of a nucleotide sequence into fragments of equal lengths (n-plets). Each whole set of n-plets, which contains $4^n$ members, is located inside one of matrices of the Kronecker family of matrices like as [A G; C T]$^{(n)}$. Correspondingly this method is closely connected with Kronecker multiplication of matrices, which is widely used in mathematics, informatics, physics, etc. and which is one of the main mathematical operation in the field of matrix genetics [Petoukhov, 2008-2013; Petoukhov, He, 2010]. Kronecker multiplication of matrices is used when one needs to go from spaces of smaller dimension into associated spaces of higher dimension. If one uses the mathematical language of vector spaces for modeling the ontogenetic complication of a living organism, it is natural to apply the ideology of a gradual transition from the spaces of low dimensions into spaces of higher dimensions. Such gradual transition is described by means of a series of Kronecker multiplication of matrices.

### 3. The description of the matrix method for long nucleotide sequences

In a general case the proposed method includes the following algorithmic steps:
1. Any long nucleotide sequence, which contains K nucleotides, is divided into equal fragments of length «n» (n-plets or n-mers), where «n» takes different values: n=1, 2, 3, …, K; in the result, an appropriate set of different symbolic representations of this sequence as a chain of n-plets appears;
2. Each n-plet in every of these representations of the sequence is transformed into three kinds of n-bit binary numbers by means of its reading from the point of view of the three sub-alphabets (Figure 2). Each of these binary numbers is transformed into its decimal equivalent. In the result, an appropriate set of different decimal representations of the initial symbolic sequence appears in a form of three kinds of sequences of decimal numbers respectively for positive integer coordinates on Cartesian axes X, Y, Z (or for numeration of rows and columns of appropriate genetic matrices).
3. Any two of the received numeric sequences define an appropriate sequence of pairs of positive integer coordinates of points on the 2-dimensional Cartesian plane (or coordinates of cells inside an appropriate genetic matrix of a Kronecker family). On the base of these pairs of coordinates, a set of corresponding points are built on the 2-dimensional Cartesian plane (or a set of corresponding cells are marked by black color inside a respective genetic matrix of a Kronecker family in contrast to other cells, which remain with white color).

In a result of these algorithmic steps, different black-and-white mosaics arise as representations of any long nucleotide sequence in different cases of its division into n-plets. Figure 3 shows examples of fractal-like and other visual patterns, which have been obtained on the base of the described method for some long nucleotide sequences.

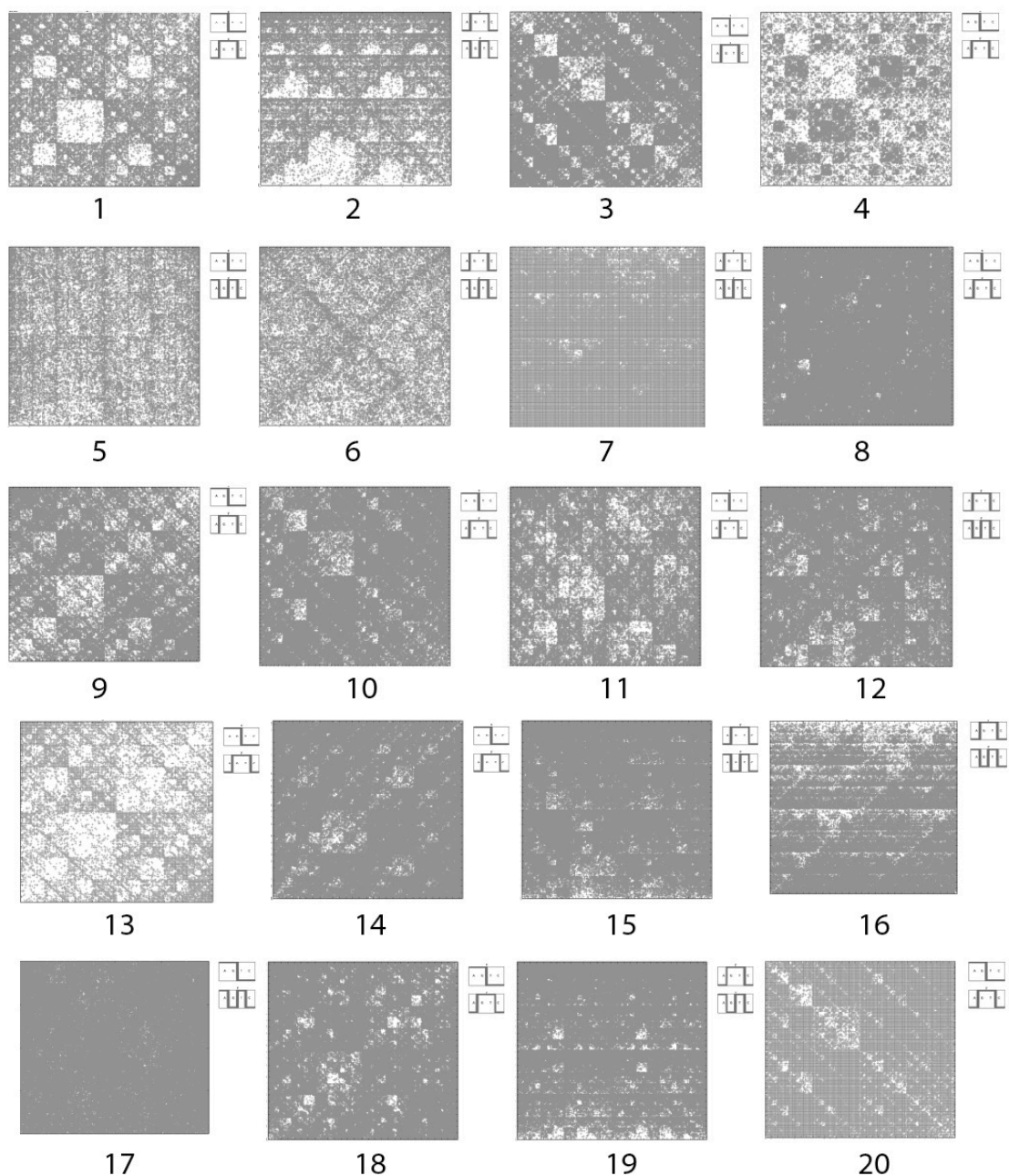

Figure 3. Examples of visual patterns, which have been received on the base of the described method for different nucleotide sequences (see explanation in the text). Two symbols are shown from the right side of each pattern to indicate what kinds of the sub-alphabets from Figure 2 were used to construct the pattern.

The numbered patterns on Figure 3 correspond to the following sequences:
1. Homo sapiens contactin associated protein-like 2 (CNTNAP2), RefSeqGene on chromosome 7 (n=63) (http://www.ncbi.nlm.nih.gov/nuccore/163954933 )
2. Homo sapiens contactin associated protein-like 2 (CNTNAP2), RefSeqGene on chromosome 7 (n=63) (http://www.ncbi.nlm.nih.gov/nuccore/163954933 )
3. Sorangium cellulosum So0157-2, complete genome (n=63) (http://www.ncbi.nlm.nih.gov/nuccore/CP003969.1 )
4. Burkholderia multivorans ATCC 17616 genomic DNA, complete genome, chromosome 2 (n=63) (http://www.ncbi.nlm.nih.gov/nuccore/AP009386.1)

5. Thermofilum sp. 1910b, complete genome (n=63)
(http://www.ncbi.nlm.nih.gov/nuccore/CP006646.1 )
6. Thermofilum sp. 1910b, complete genome (n=63)
(http://www.ncbi.nlm.nih.gov/nuccore/CP006646.1)
7. Dinoroseobacter shibae DFL 12, complete genome (n=8)
(http://www.ncbi.nlm.nih.gov/nuccore/CP000830.1)
8. Escherichia coli LY180, complete genome (n=24)
(http://www.ncbi.nlm.nih.gov/nuccore/NC_022364.1)
9. Francisella tularensis subsp. tularensis SCHU S4 complete genome (n=24)
(http://www.ncbi.nlm.nih.gov/nuccore/AJ749949.2)
10. Halomonas elongata DSM 2581, complete genome (n=24)
(http://www.ncbi.nlm.nih.gov/nuccore/FN869568.1)
11. Helicobacter mustelae 12198 complete genome (n=24)
(http://www.ncbi.nlm.nih.gov/nuccore/FN555004.1)
12. Helicobacter mustelae 12198 complete genome (n=12)
(http://www.ncbi.nlm.nih.gov/nuccore/FN555004.1)
13. Invertebrate iridovirus 22 complete genome (n=8)
(http://www.ncbi.nlm.nih.gov/nuccore/NC_021901.1)
14. Methanosalsum zhilinae DSM 4017, complete genome (n=12)
(http://www.ncbi.nlm.nih.gov/nuccore/CP002101.1)
15. Methanosalsum zhilinae DSM 4017, complete genome (n=12)
(http://www.ncbi.nlm.nih.gov/nuccore/CP002101.1)
16. Mycobacterium abscessus subsp. bolletii INCQS 00594
INCQS00594_scaffold1, whole genome shotgun sequence (n=12)
(http://www.ncbi.nlm.nih.gov/nuccore/544224292)
17. Penicillium chrysogenum Wisconsin 54-1255 complete genome, contig
Pc00c12 (n=32) (http://www.ncbi.nlm.nih.gov/nuccore/AM920427.1)
18. Riemerella anatipestifer DSM 15868, complete genome (n=12)
(http://www.ncbi.nlm.nih.gov/nuccore/CP002346.1)
19. Riemerella anatipestifer DSM 15868, complete genome (n=12)
(http://www.ncbi.nlm.nih.gov/nuccore/CP002346.1)
20. Burkholderia multivorans ATCC 17616 genomic DNA, complete genome,
chromosome 2 (n=8) (http://www.ncbi.nlm.nih.gov/nuccore/AP009386.1)

Such mosaic pattern shows phenomenology of «presence-and-absence» of different n-plets. One should note that a division of a long nucleotide sequence only into a single of possible variants of its equal fragmentation (for example, a division into 16-plets) doesn't give an unambiguous definition of this sequence: such separate case of a division represents this sequence as a set of fragments but without a reflection of their order in the sequence (any permutation of these fragments gives a new sequence with the same set of n-plets). To get an unambiguous definition of the sequence, one should take into consideration all (or many) possible variants of its equal partitions (n = 1, 2, 3,….). In practice for many tasks of a comparison analysis and classification of different long nucleotide sequences it is enough to consider some chosen variants of fragmentations of these sequences, for example, variants with n = 16, 32, 64.

Another possible way of an unambiguous representation of a long nucleotide sequence in a case of its division with a certain value n (for example, with n=8) is connected with a construction of additional visual patterns, which reflect an order of n-plets in the sequence.

Figure 4 shows two examples of such mosaic patterns for Homo sapiens chromosome 22 genomic scaffold and for Arabidopsis thaliana mitochondrion in the case of their representations as sets of 16-mers. On these mosaics, white places correspond to dispositions of those 16-mers on a corresponding 2-dimensional plane, which are missing in such representations of the sequences. The mosaic pattern depends on a concrete choice of two kinds of sub-alphabets from Figure 2. Figure 4 shows two mosaic patterns on 2-dimensional Cartesian planes (x,y) and (y,z), which are identical to black-and-white mosaics of the genetic matrices [A G; C T][16] and [G T; C A][16] respectively, where cells with existing 16-plets of the sequence are shown by black color and cells with missing 16-plets are shown by white color.

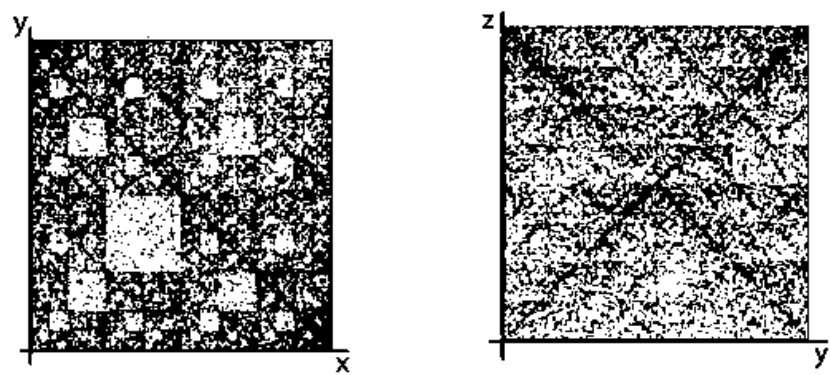

Figure 4. Two examples of patterns which are constructed on the base of the described method. Left side: the visual pattern of the nucleotide sequence Homo sapiens chromosome 22 genomic scaffold, which has 648059 nucleotides (http://www.ncbi.nlm.nih.gov/nuccore/NW_004078110.1?report=genbank) and which is divided into a sequence of 16-mers; these 16-mers are transformed into 16-bit binary numbers on the basis of the first sub-alphabet and of the second sub-alphabet (Figure 2); then their decimal equivalents are postponed on the axes "x" and "y" respectively. Right side: the visual pattern of the nucleotide sequence Arabidopsis thaliana mitochondrion, which has 366924 nucleotides (http://www.ncbi.nlm.nih.gov/nuccore/NC_001284.2) and which is divided into a sequence of 16-mers; these 16-mers are transformed into 16-bit binary numbers on the basis of the second sub-alphabet and of the third sub-alphabet (Figure 2); then their decimal equivalents are postponed on the axes "y" and "z" respectively.

Figure 5 shows one of interesting patterns received by the described method.

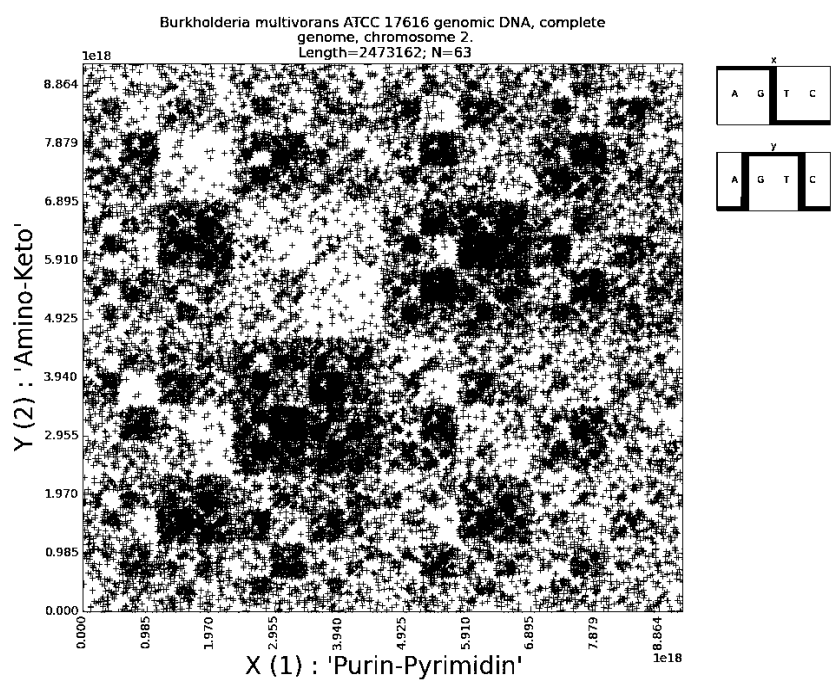

Figure 5. The pattern of the sequence "Burkholderia multivorans ATCC 17616 genomic DNA, complete genome, chromosome 2" with 2473162 nucleotides (http://www.ncbi.nlm.nih.gov/nuccore/AP009386.1) in the case of its division into 63-plets.

Binary representations of n-mers are expressed in a form of n-bit binary numbers, the quantity of kinds of which is equal to $2^n$. For example, the set of 3-bit binary numbers contains $2^3=8$ members: 000, 001, 010, 011, 100, 101, 110, 111 (their equivalents in decimal notation are 0, 1, 2, 3, 4, 5, 6, 7). Decimal equivalent of the biggest n-bit binary member in a set of n-bit binary numbers is equal to $2^n-1$. Such sets of n-bit binary numbers are named "dyadic groups" (see details in [Petoukhov, 2013]).

The most interesting application of this matrix method is realized for a case of long nucleotide sequences, which are divided into relative long n-mers (n=8, 9, 10, …). Reasons for this are the following (see Figure 6):

- a long nucleotide sequence, which is divided into relative short n-mers (n=1, 2, 3, 4), contains usually all possible kinds of such short n-mers; correspondingly its visual pattern is trivial because it contains all possible points with positive integer coordinates (x,y) inside an appropriate numeric range;
- a long nucleotide sequence, which is divided into relative long n-mers (n=8, 9, 10, …), usually generates a regular non-trivial mosaic of a fractal-like or other character. This was detected using a special computer program in a course of initial investigations of different long nucleotide sequences by means of the described method.

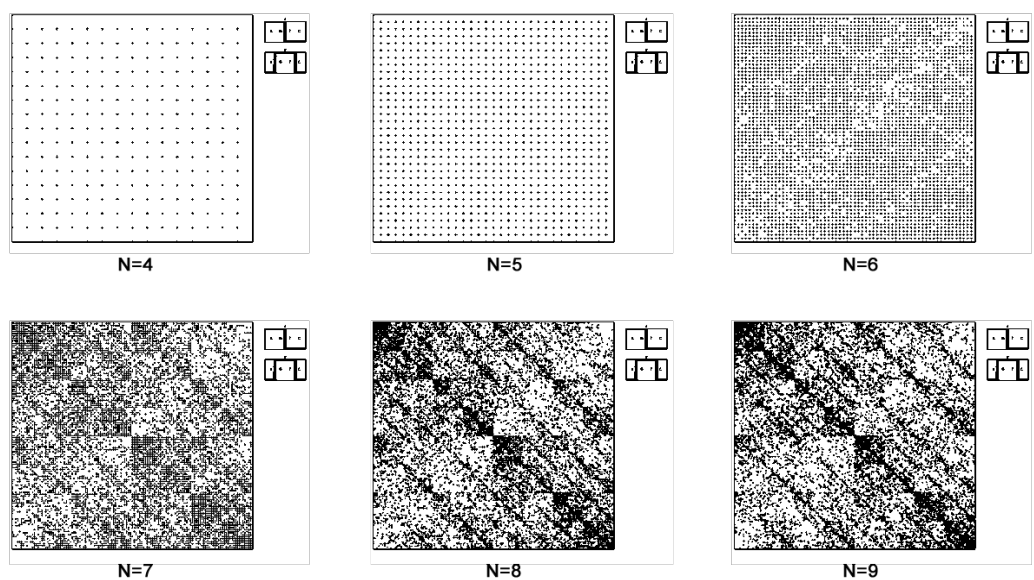

Figure 6. Examples of patterns for the sequence Fistulifera sp. JPCC DA0580 chloroplast, complete genome (http://www.ncbi.nlm.nih.gov/nuccore/330850815 ) in cases of its divisions into n-plets with n = 4, 5, 6, 7, 8, 9.

Figure 6 (lower level) also illustrates that - in a certain range of changing values «n» -  visual fractal mosaics for different «n» are approximately repeated each other (see Section 5 about this «stability» of the fractal-like mosaics).

Fractal patterns, which are obtained by means of the described matrix method, sometimes resemble fractal patterns of long nucleotide sequences and amino acid sequences, which were previously obtained by means of the known method "Chaos Game Representation" (CGR-method) in works [Jeffrey, 1990; Petoukhov, He, 2010; etc.] though both methods are quite different in their algorithmic essence. In particularly, CGR-method deals with representations of nucleotide sequences or other long sequences by means of four numbers 0, 1, 2, 3 but not by means of binary numbers 0, 1. In addition our new method seems to be more simple for understanding and using by biologists.

## 4. Long random sequences

What kinds of visual patterns are produced by the described method in cases of long random sequences of nucleotides? To answer on this question, different random sequences were generated by a computer program. Their study is revealed that appropriate visual patterns have non-regular characters in contrast to cases of real genomes. Figure 7 shows examples of visual patterns for a case of the random sequence with 100000 nucleotides in cases of its division into n-plets with n = 8, 16, 28 (this sequence is disposed at website pentagramon.com for its possible additional testing). Each of visual patterns of this random sequence for two other 2-dimensional planes (x,z) and (y,z) has a similar non-regular character.

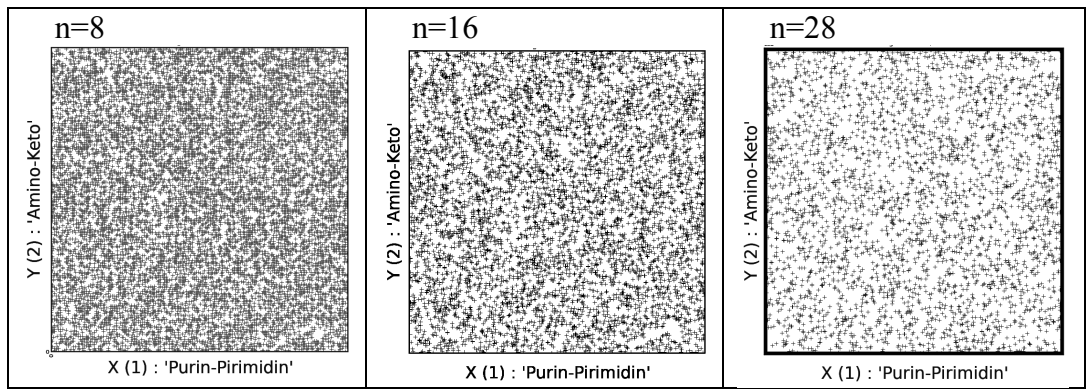

Figure 7. Examples of visual patterns of a random nucleotide sequence with 100000 nucleotides (pentagramon.com) in cases of its division with n = 8, 16, 28.

## 5. Kronecker multiplication, fractal lattices and the problem of coding an organism on different stages of its onthogenesis

Previous Sections have shown that the described method gives very different types of visual patterns for random nucleotide sequences (where non-regular patterns arise as on Figure 7) and for real nucleotide sequences (where fractal-like patterns arise frequently as on Figures 3-6). The authors note that in many cases these fractal-like patterns of long nucleotide sequences resemble fractal lattices, which are automatically generated for matrices of Kronecker families. One should explain this in more details.

Let us take a square (k*k)-matrix M, whose entries are equal only to 0 or 1. Any integer Kronecker power (n) of this matrix generates a new $(k^n*k^n)$-matrix $M^{(n)}$ with a fractal location of entries 0 and 1 inside it (Figures 8, 9). These fractal mosaics inside such matrices of Kronecker families are called "fractal lattices". The theme "Kronecker multiplication and fractal lattices" are accurately described in the book [Gazale, 1999, Section X]. Such fractal lattices (Figure 8) are generated due to a general definition of Kronecker multiplication of matrices as a special mathematical operation.

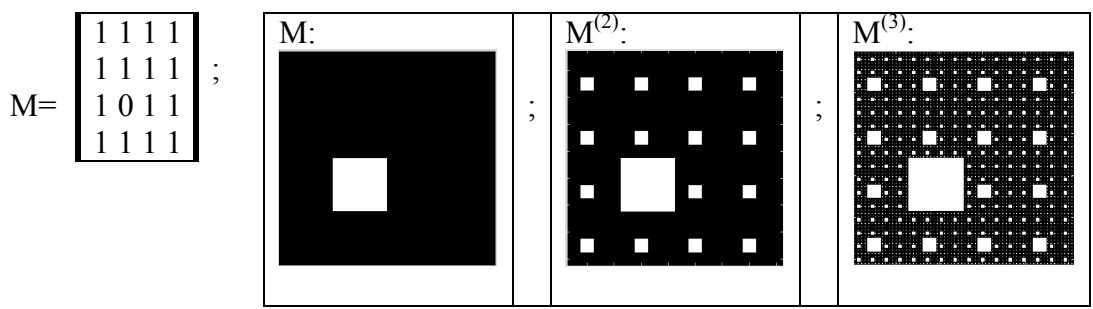

Figure 8. An example of generating fractal lattices by means of Kronecker exponentiation of matrices. Left side: the (4*4)-matrix M with entries 0 and 1. Right side: visu-

al patterns of the matrix M and its Kronecker powers $M^{(2)}$ and $M^{(3)}$, which are (16*16)-matrix and (64*64)-matrix respectively. Here black color corresponds to matrix cells with entries 1 and white color corresponds to cells with entries 0.

One should note that, in many cases, significant features of fractal-like patterns of real nucleotide sequences can be simulated by means of fractal lattices of matrices of a Kronecker family, if a matrix kernel of the Kronecker family is adequately chosen. For example let us take the pattern (from Figure 4) of the nucleotide sequence Homo sapiens chromosome 22 genomic scaffold, which has 648059 nucleotides (http://www.ncbi.nlm.nih.gov/nuccore/NW_004078110.1?report=genbank) and which is divided into a sequence of 16-mers. If this pattern is covered by the uniform (8*8)-grid, 8 cells of this grid will be almost white color in contrast to the remaining 56 cells (Figure 9, upper level, left side). In such case this (8*8)-mosaic of black-and-white type is identical to mosaic of the genetic (8*8)-matrix $[A\ G;\ C\ T]^{(3)}$ of 64 triplets where those 8 triplets are missing, which are located in this matrix on the same places and which are marked by red color on Figure 9 (upper level, right side). Let us replace these 8 missing triplets by number 0, and all other 56 triplets by number 1. It leads to a transformation of this variant of symbolic matrix $[A\ G;\ C\ T]^{(3)}$ into a numeric matrix S (Figure 9, bottom level, left side).

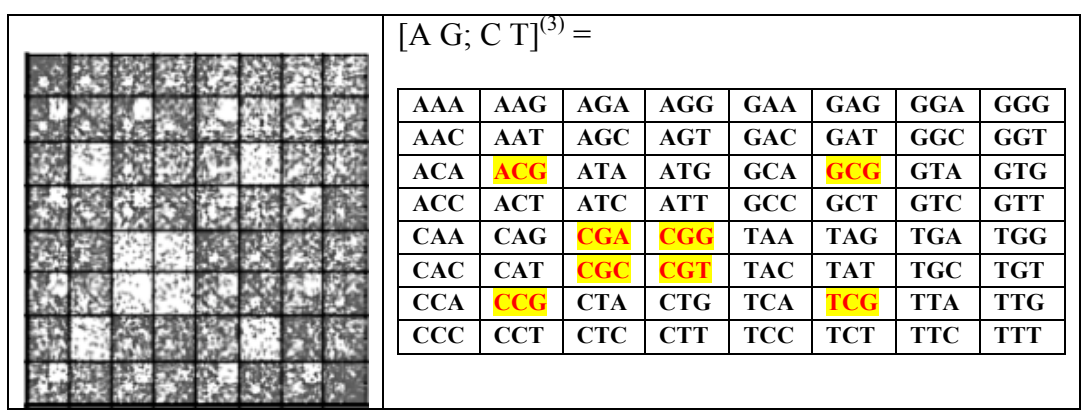

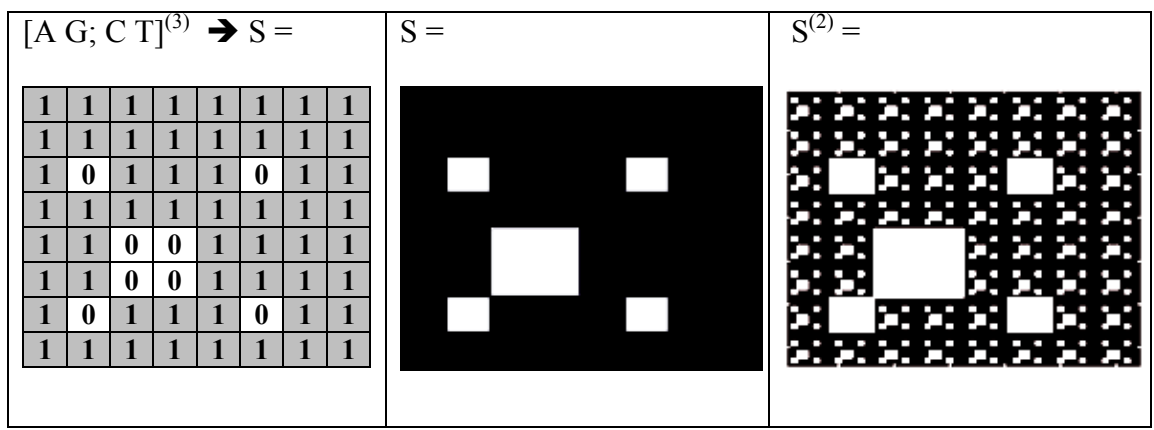

Figure 9. Illustration to relations among Kronecker multiplication, fractal lattices and fractal-like patterns of long nucleotide sequences (explanations in text).

Kronecker exponentiation of the matrix S generates matrices $S^{(2)}$, $S^{(3)}$,..., whose visual patterns demonstrate appropriate fractal lattices, one of which for the matrix $S^{(2)}$ is shown on Figure 9 (bottom level, right side). The numeric matrix $S^{(16)}$ contains the whole set of 16-plets with an appropriate fractal lattice, which resembles the visual pattern of the real nucleotide sequence Homo sapiens chromosome 22 genomic scaffold on Figure 4. One should note that the visual pattern of this real sequence contains more white places (than in the matrix $S^{(16)}$) because many additional 16-plets are absent since the sequence has a finite length in 648059 nucleotides.

Fractal-like lattices in visual patterns of long nucleotide sequences testify in favor of significance of Kronecker multiplication for structuration of these genetic sequences. This is not an isolated fact about a genetic significance of Kronecker multiplication. Previously we have received other evidences about a biological significance of Kronecker multiplication of matrices in phenomenology of natural ensembles of molecular-genetic alphabets [Petoukhov, 2008, 2012a,b, 2013a,b; Petoukhov, He, 2010] and also in a structure of Punnet squares in the field of Mendelian genetics in a connection with the Mendelian laws of independent inheritance of traits [Petoukhov, 2011].

What possible reasons of the genetic significance of Kronecker multiplication can be named? One of main tasks of the genetic encoding system of an organism is that this system should encode a combined ensemble of biostructures which becomes more and more complex in a course of onthogenesis. Onthogenesis of an organism is a process of receiving a great number of new degrees of freedom for the organism.

From mathematical point of view, it means that internal space of the organism becomes more and more n-dimensional («n» increases step by step in a course of onthogenesis). It is obvious that these are very different things: encoding the embryonic body, and encoding the adult organism, which has grown from it. Mathematics of genetic coding should correspond to this ontogenetic development of the organism with increasing organismic degrees of freedom, which means an increasing dimension of its internal vector space. (A similar conception of an increasing n-dimension of internal spaces can be also applied to phylogenetic increasing complexity of organisms in biological evolution). Respectively such mathematics of multidimensional spaces should model appropriate facts of onthogenetic and phylogenetic increasing complexity of organisms. This aspect seems to be one of the main reasons of biological significance of Kronecker multiplication. The increasing complexity and evolution of the genetic system itself can take place on similar principles: they are also associated with increasing n-dimensions of the corresponding genetic spaces, which can be simulated by means of using Kronecker multiplication. We suppose that the connection of the genetic system with Kronecker multiplication reflects the ability of this system to encoding inherited structures inside associated multi-dimensional internal spaces of developing organisms. The authors suppose that it is impossible to understand the system of genetic coding enough deeply without taking into account the fundamental task of the genetic encoding inside an individual set of multi-dimensional spaces, which accompanies onthogenetic development of organism.

### 6. Patterns of human chromosomes

What kinds of binary mosaics are generated by means of the described method for all 23 pairs of human chromosomes, data of which can be taken from website ftp://ftp.ncbi.nih.gov//genomes/H_sapiens/April_14_2003/? Our results of their testing testify that they are represented by binary mosaics of analogical types. Figure 10 shows mosaics of the first 15000000 (fifteen millions) nucleotides of the following sequences in the case of their division into 63-plets: Homo sapiens chromosomes X and Y together with Homo sapiens chromosome 1 (they have 152634166, 50961097 and 245203898 nucleotides respectively).

gi|29826146|ref|NC_000023.4|NC_000023 Homo sapiens chromosome X, complete sequence
Length=152634166; N=63

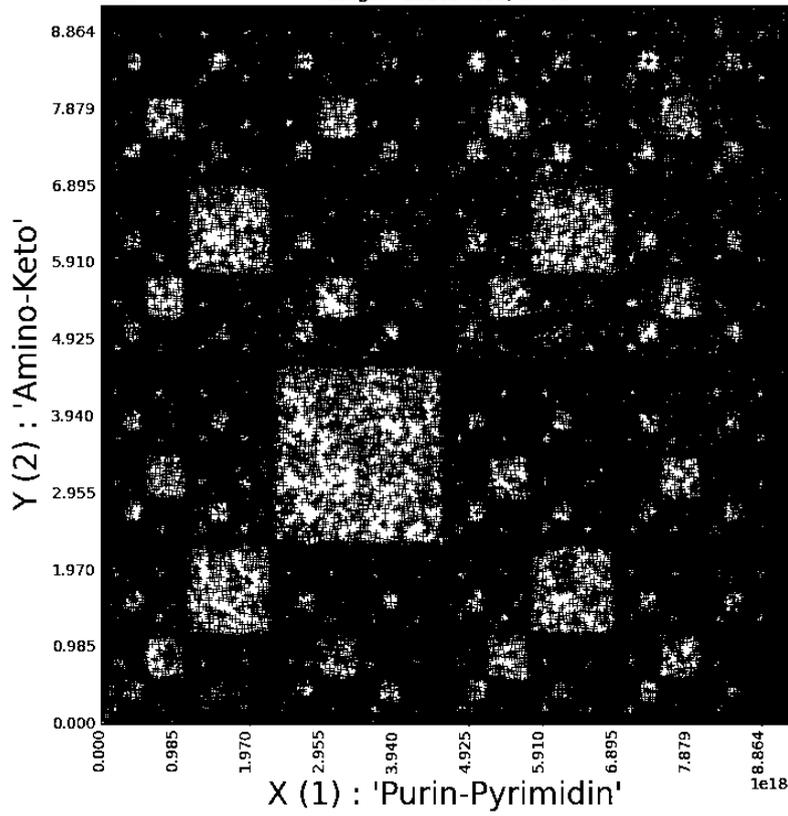
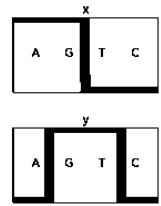

gi|29824594|ref|NC_000024.3|NC_000024 Homo sapiens chromosome Y, complete sequence
Length=50961097; N=63

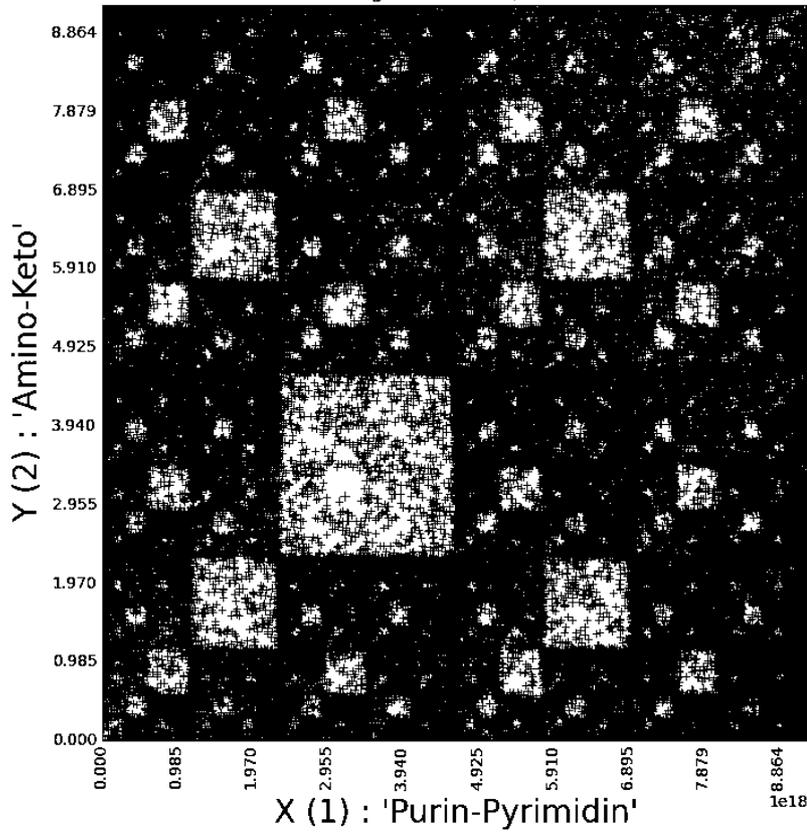
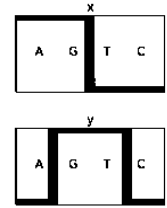

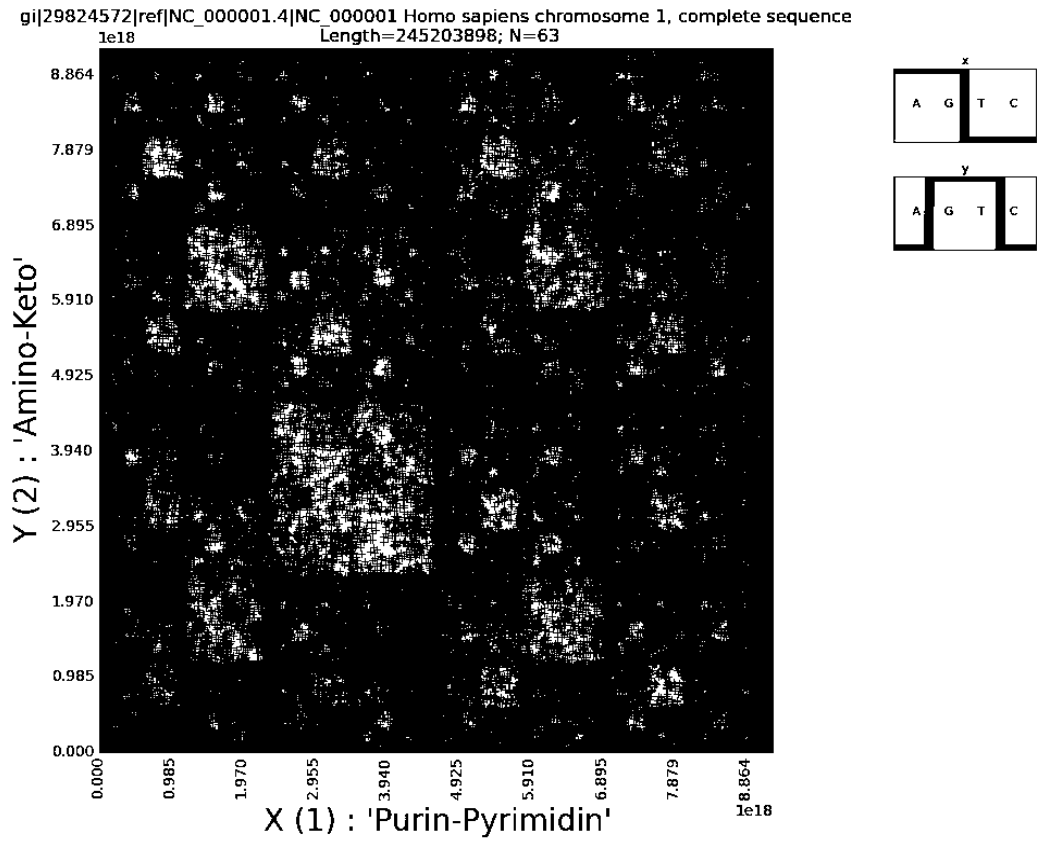

Figure 10. The binary mosaics of the first 15000000 nucleotides of the sequences of Homo sapiens chromosomes X and Y (two upper levels) and Homo sapiens chromosome 1 in the case of their division into 63-plets. Initial data were taken from website ftp://ftp.ncbi.nih.gov//genomes/H_sapiens/April_14_2003/

     Also we tested the first 15000000 (fifteen millions) nucleotides of every human chromosome from the mentioned website. Structures of 2-dimensional mosaics of these tested sequences were very similar.

     Then we took different parts of the same sequence Homo sapiens chromosome 1; every of parts had 15000000 (fifteen millions) nucleotides. Again our results of their testing testify that these parts generate very similar mosaics by the described method. Figure 11 shows examples of two parts of this long sequences: one part corresponds to interval of this sequence from 45000000 to 60000000 nucleotides, and the second part corresponds to interval from 135000000 to 150000000 nucleotides.

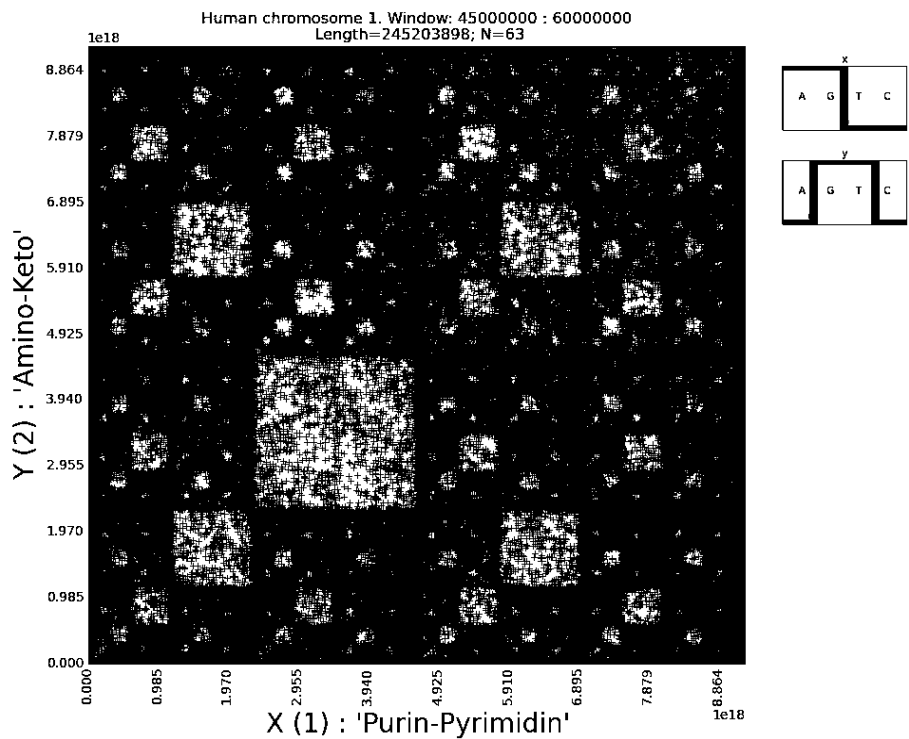

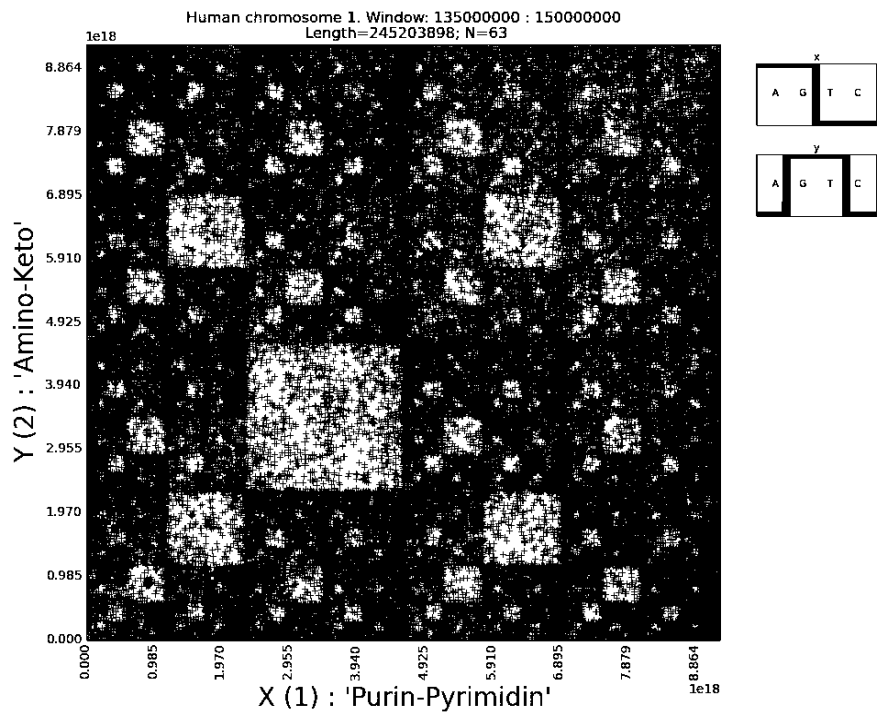

Figure 11. Examples of binary mosaics for two parts of Homo sapiens chromosome 1 (ftp://ftp.ncbi.nih.gov//genomes/H_sapiens/April_14_2003/), which has 245203898 nucleotides: one part corresponds to interval of this sequence from 45000000 to 60000000 nucleotides, and the second part corresponds to interval from 135000000 to 150000000 nucleotides

### 7. Patterns of penicillins

Additionally we have tested different kinds of penicillins by the described method to receive their binary patterns. The results testify that in this group of antibiotics their long nucleotide sequences usually generate non-regular mosaics, which resemble mosaics of random nucleotide sequences. Perhaps important medical meaning of penicillins is connected with this their peculiarity.

Figure 12 shows some examples of the binary mosaics for different contigs of Penicillium chrysogenum Wisconsin 54-1255 complete genome.

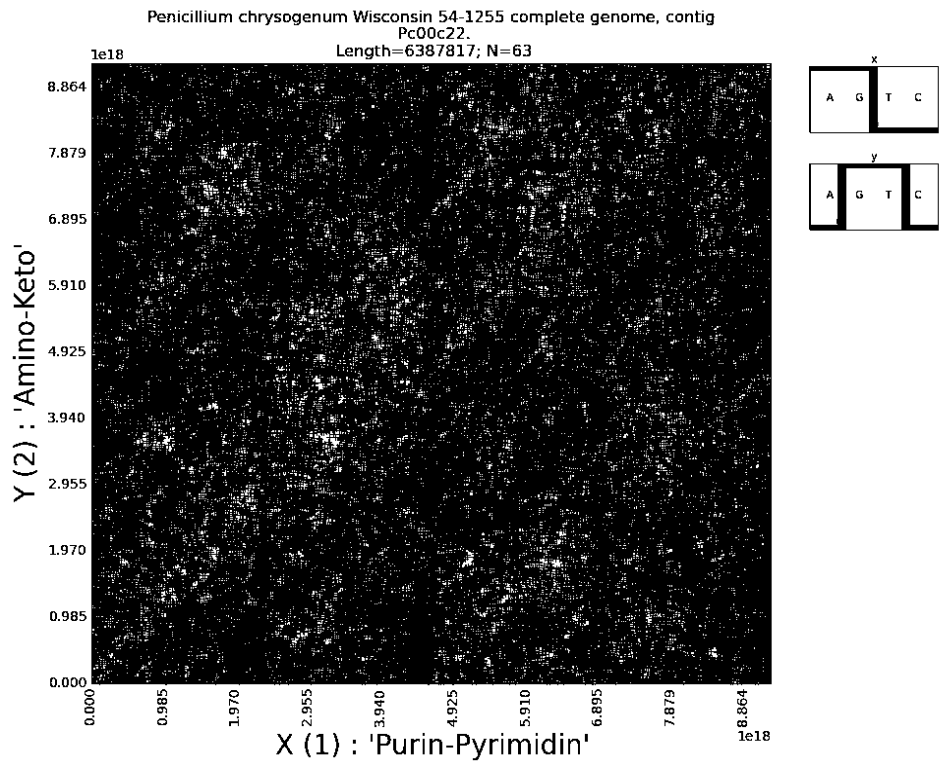

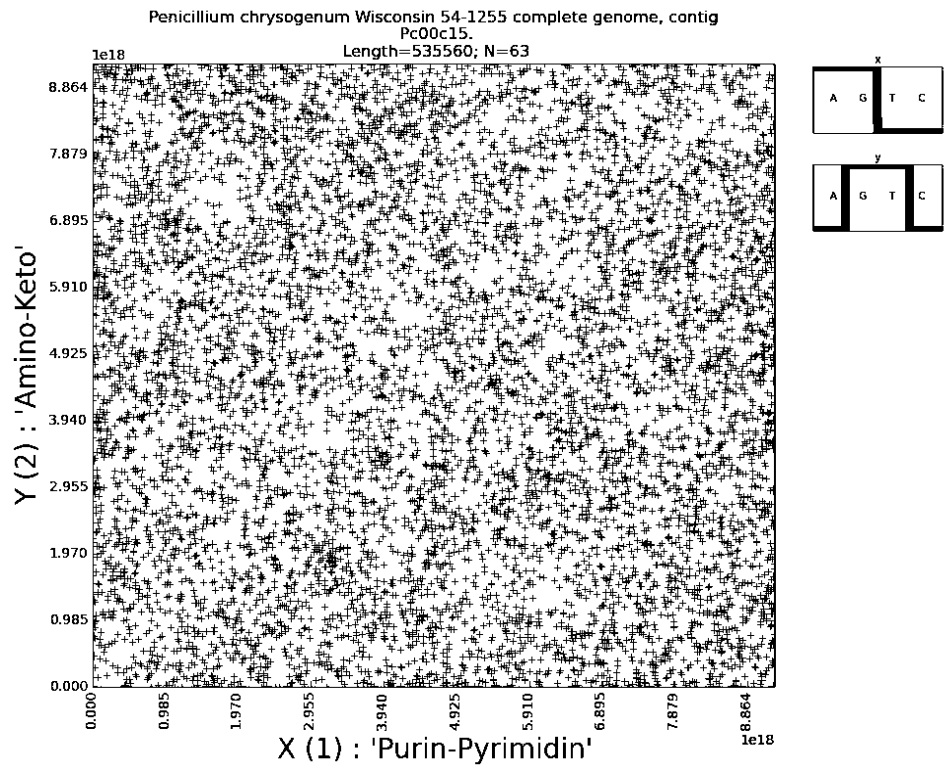

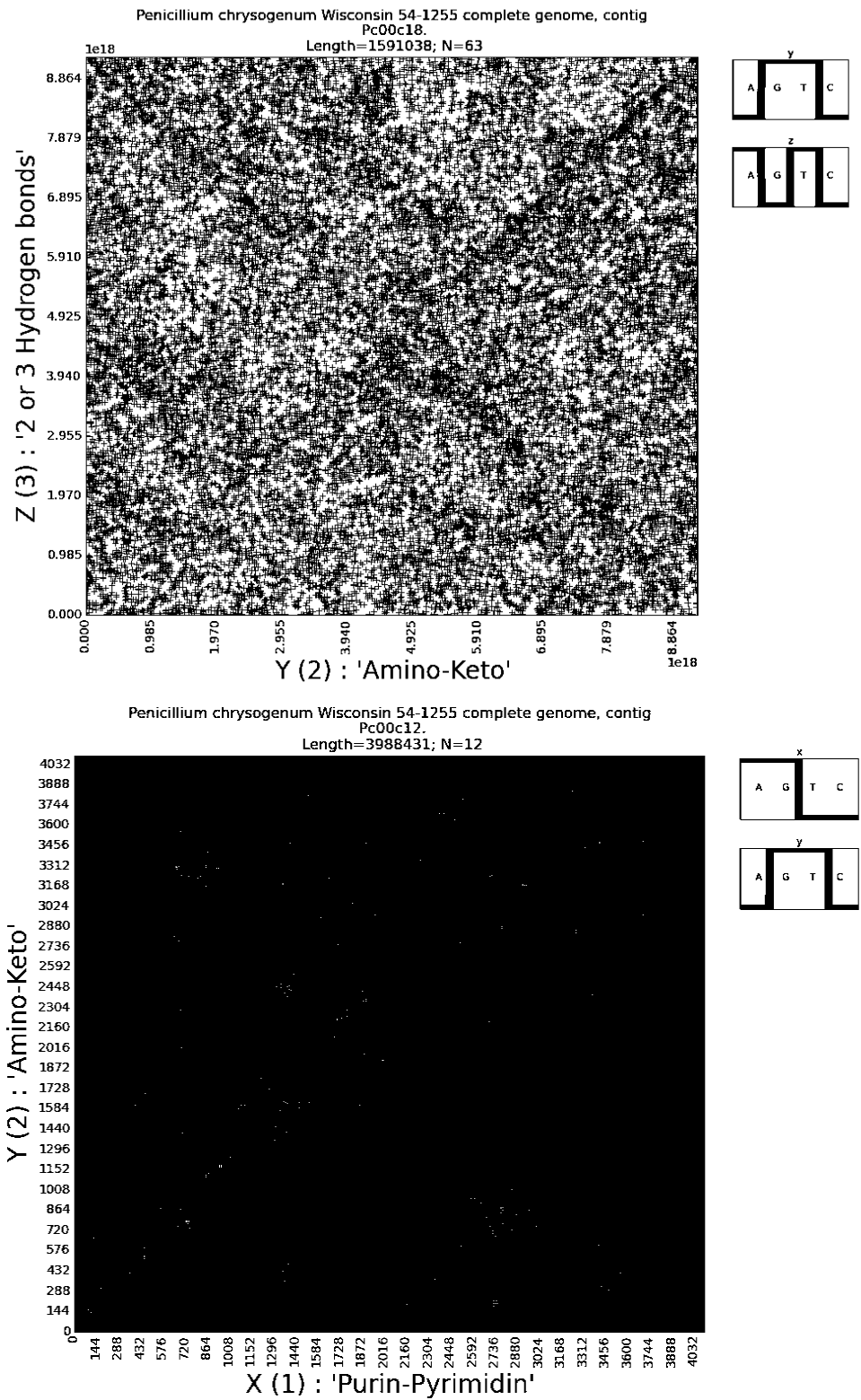

Figure 12. Examples of binary mosaics for long nucleotide sequences of different contigs of Penicillium chrysogenum Wisconsin 54-1255 complete genome (http://biocyc.org/PCHR/organism-summary?object=PCHR). Upper level: the mosaic for the contig 22, which contains 6387817 nucleotides, for the case of 63-plets. The second level: the mosaic for the contig 15, which contains 535560 nucleotides, for the case of 63-plets. The third level: the mosaic for the contig 18, which contains 1591038 nucleotides, for the case of 63-plets. Lower level: the mosaic for the contig 12, which contains 3988431 nucleotides, for the case of 12-plets. Symbols from the right side of each mosaic indicate the pair of the sub-alphabets that were used for a transformation of these n-plets into binary numbers.

## 8. About 3D-representations

Till now we told about 2-dimensional representations of long nucleotide sequences by means of the described method. But it is obvious that 3d-patterns can be also constructed in a similar way on basis of all the 3 binary (and decimal) representations of any nucleotide sequence by means of the three sub-alphabets from Figure 2.

One can initially consider a special case when on 2-dimensional Cartesian plane (x,y) all points with positive integer coordinates exist in the range of coordinate decimal values from (0, 0) till (100, 100) or in the range of their binary values from (0, 0) till (1100100, 1100100). Above in the Section 2 it was noted that the binary representation of any nucleotide n-plet from the point of view of the third sub-alphabet (Figure 2) can be received by means of modulo-2 addition of its two binary representations from the points of view of the first and second sub-alphabets. Correspondingly we suppose that a value of coordinate «z» of every of considered points (x,y) is defined as a sum of its binary coordinates «x» and «y» on base of modulo-2 addition. A set of points (x,y,z) of an «ideal» 3-dimensional configuration arises in this case. This ideal 3d-configuration has a non-simple character and contains all possible points of the considered range (or all corresponding n-plets).

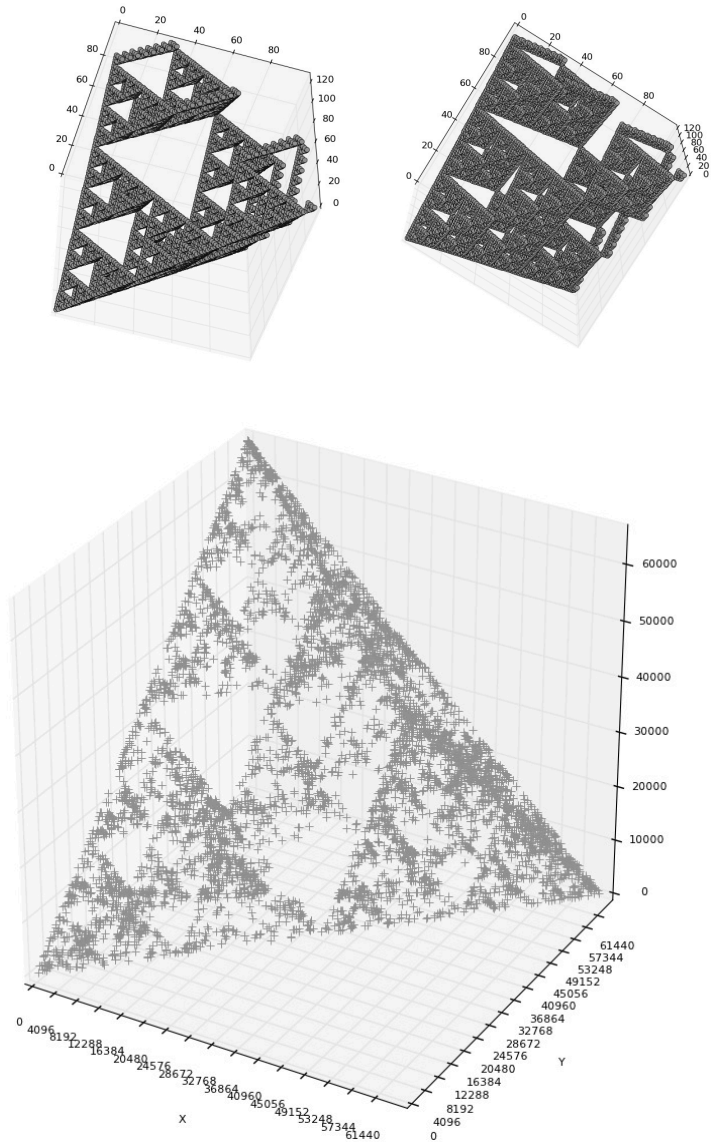

Figure 13. Upper level: two 2-dimensional images of an «ideal» 3d-configuration in its examination from two oblique foreshortenings. Lower level: a real 3d-configuration for a real nucleotide sequence (see explanation in text).

Figure 13 (upper level) shows two 2-dimensional images of this ideal 3d-configuration in its examination from two oblique foreshortenings. Also Figure 13 (lower level) shows a real 3d-configuration, which was constructed for a real nucleotide sequence by means of the same definition of its third coordinate «z» (as a sum of binary values of coordinates «x» and «y» of n-plets of the sequence). This real 3d-configuration differs from the ideal 3d-configuration due to many additional «white» areas in its structure because of absence of appropriate number of n-plets. Projections of the real 3d-configuration of a nucleotide sequence into 2-dimensional planes (x,y), (x,z) or (y,z) give the corresponding 2d-patterns of the sequence.

## 9. Gray code and binary representations of long nucleotide sequences

It is known that Gray code (http://en.wikipedia.org/wiki/Gray_code), which is widely explored to facilitate error correction in digital communications, uses a special system of relation between binary numbers and decimal numbers. Some questions about a possible connection between genetic matrices and Gray code were described in the work [Kappraff, Petoukhov, 2009]. One can mention about a connection of Gray code with QR codes (http://en.wikipedia.org/wiki/QR_code). Now we have tested a possible interpretation of binary presentations of long nucleotide sequences from the point of view of Gray code (in contrast to using the conventional binary code in the previous Sections of this article).

This testing testify that in the case of using Gray code final mosaic patterns of long nucleotide sequences don't usually demonstrate such expressive regularities as in the case of using the conventional binary code. Figure 14 shows examples of binary mosaics, which were received in the case of using Gray code by means of the described method.

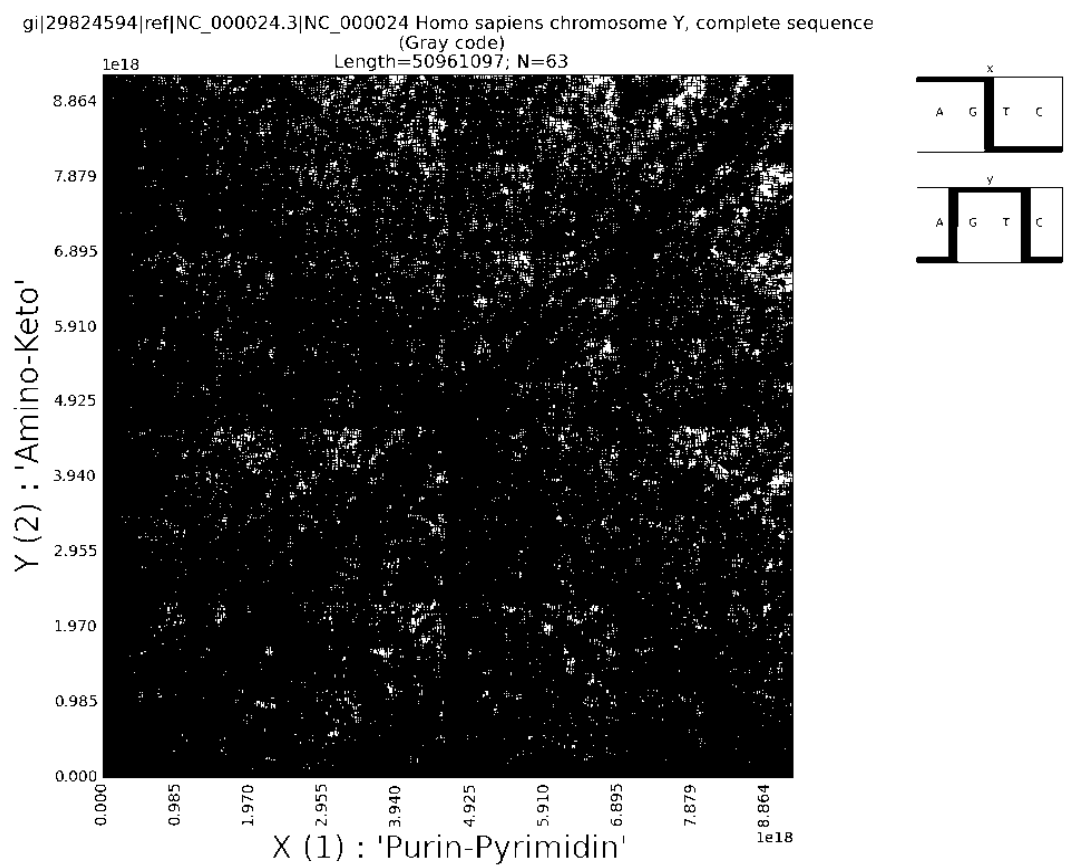

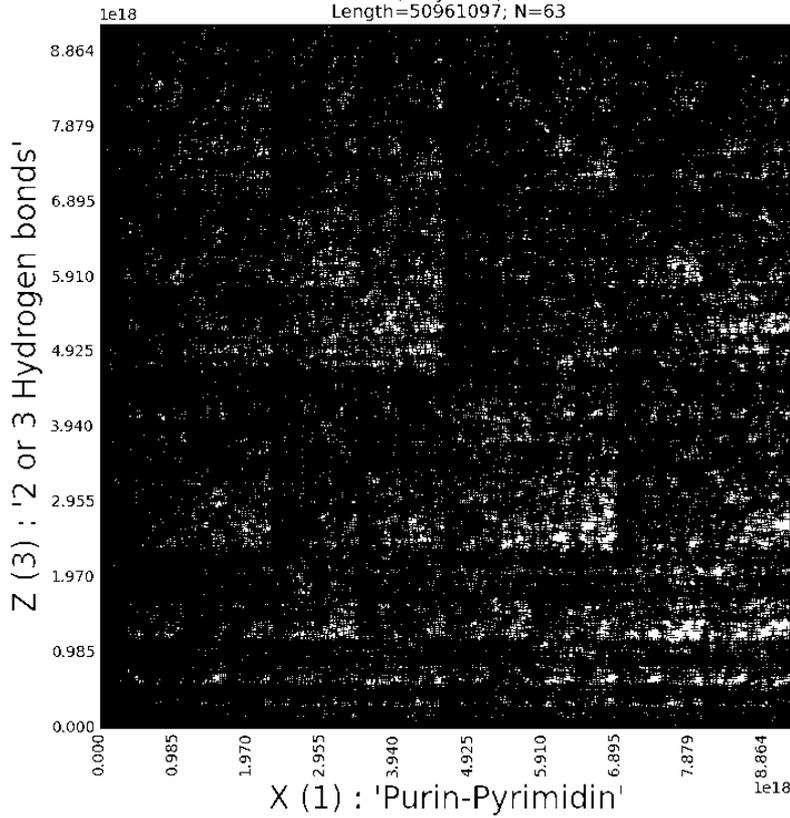
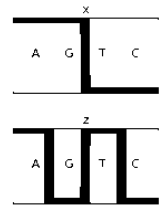

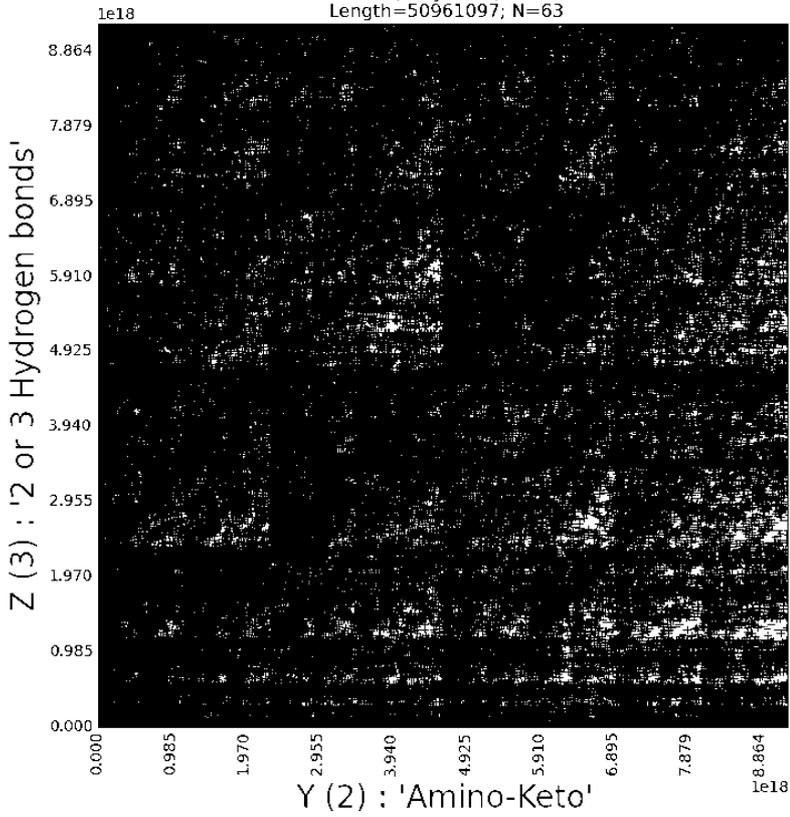
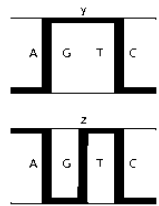

Figure 14. Binary mosaics in the case of using Gray code for the first 15000000 nucleotides of the sequence Homo sapiens chromosome Y (ftp://ftp.ncbi.nih.gov//genomes/H_sapiens/April_14_2003/), which was divided into 63-plets. Symbols from the right side of each mosaic indicate the pair of the sub-alphabets that were used for a transformation of these 63-plets into binary numbers

## 10. Some concluding remarks

We have obtained the first results of the application of this new method of analysis of long nucleotide sequences. The preliminary results include, for example, the stability of resulting fractal-like and other patterns in a case of shifts of reading frame of such sequences, or in a case of reversing of sequences, or in a case of the permutation of fragments of a sequence, or in case of a removal of certain parts of sequences; these results are mainly similar to the results of studies of fractal genetic networks for long nucleotide sequences [Petoukhov, Svirin, 2012]. In particularly, we saw a stability of mosaic patterns in cases of transformations of examined nucleotide sequences by means of removal of the every second nucleotide in sequences, or removal of the every third nucleotide in sequences, etc. Many adjacent variants can be added to the described method for deeper research of long genetic sequences by means of their binary presentation (for example, quantities of elements 0 and 1, which are met in n-plets of two kinds of the n-bit binary presentation of a long nucleotide sequence, can be used to construct a new type of a visual pattern of the sequence).

Different types of genetic sequences (for example amino acid sequences) can be also represented in a form of pairs of binary sequences through the use of various sets of their binary-oppositional attributes; in these cases, this method of analysis of their content can also be applied.

The described method of analysis of long nucleotide sequences seems to be useful for the study of hidden regularities in genetic sequences and also for classification and comparative analysis of different genetic sequences with possible applications in biotechnology and medicine. This article adds new materials into the field of algebraic biology, where matrix methods seem to be extremely useful [Petoukhov, 2008; 2012a,b; 2013a,b; Petoukhov, He, 2010]. The discovery of new binary fractal-like patterns, which are revealed from long genetic sequences by means of this method, provokes many questions about relations between the genetic system and those fields of science and technology, where digital binary fractals are used, for example, fields of fractals in radiophysics, technology of fractal antenna, fractal codes, etc.

Computer software for applications of this method was created by I.Stepanian. Many of the described results were received by means of using the supercomputer of "the Joint Supercomputer Center of the Russian Academy of Sciences" (http://www.jscc.ru/eng/index.shtml).